\documentclass[journal]{IEEEtran}
\usepackage{amsmath,amsfonts}
\usepackage{algorithmic}
\usepackage{algorithm}
\usepackage{array}
\usepackage{textcomp}
\usepackage{stfloats}
\usepackage{verbatim}
\usepackage{cite,url,subfigure,epsfig,graphicx}
\IEEEoverridecommandlockouts
\makeatletter
\def\ps@headings{
\def\@oddhead{\mbox{}\scriptsize\rightmark \hfil \thepage}
\def\@evenhead{\scriptsize\thepage \hfil \leftmark\mbox{}}
\def\@oddfoot{}
\def\@evenfoot{}}
\makeatother 

\usepackage{times,verbatim,amsfonts,amsmath,color}
\begin{document}

\title{Social Metaverse: Challenges and Solutions}
\author{Yuntao~Wang, Zhou~Su, and Miao~Yan
\thanks{Manuscript accepted May 23, 2023 by IEEE Internet of Things Magazine.}
\thanks{Y. Wang, Y. Pan, M. Yan, Z. Su, and T. H. Luan are with the School of Cyber Science and Engineering, Xi'an Jiaotong University, Xi'an, China (\emph{Corresponding author: Zhou~Su}).}}

\maketitle

\begin{abstract}
Social metaverse is a shared digital space combining a series of interconnected virtual worlds for users to play, shop, work, and socialize. In parallel with the advances of artificial intelligence (AI) and growing awareness of data privacy concerns, federated learning (FL) is promoted as a paradigm shift towards privacy-preserving AI-empowered social metaverse.
However, challenges including privacy-utility tradeoff, learning reliability, and AI model thefts hinder the deployment of FL in real metaverse applications.
In this paper, we exploit the pervasive social ties among users/avatars to advance a social-aware hierarchical FL framework, i.e., \emph{SocialFL} for a better privacy-utility tradeoff in the social metaverse. Then, an aggregator-free robust FL mechanism based on blockchain is devised with a new block structure and an improved consensus protocol featured with on/off-chain collaboration.
Furthermore, based on smart contracts and digital watermarks, an automatic federated AI (FedAI) model ownership provenance mechanism is designed to prevent AI model thefts and collusive avatars in social metaverse.
Experimental findings validate the feasibility and effectiveness of proposed framework.
Finally, we envision promising future research directions in this emerging area.
\end{abstract}


\section{Introduction}
\IEEEPARstart{M}{etaverse} is a collectively shared, fully immersive, hyper spatiotemporal, and physically persistent 3D virtual space, which is parallel yet interactive to the real world \cite{ref1}. 
Through any type of smart device from smart phones to head-mounted displays, human beings can freely enter the metaverse realm to communicate, collaborate, entertain and socialize with other avatars and digital things, as well as experience an alternate life.
The metaverse is generally regarded as the future iteration of the Internet. Instead of ``browsing'' information in the current mobile Internet, users ``live'' in the metaverse as avatars to immersively make digital creations and freely share them across various sub-metaverses.
Due to the intrinsic openness and decentralization features, metaverse will inevitably inspire new innovations for social applications and reshape the concepts of social interaction, social experience and social networks, making the social metaverse becoming a platform of such new social ecosystem.

In social metaverse, by disassembling entrenched social platforms and dismantling siloed social applications, it allows massive and intensive social interactions among user-controlled avatars, as well as immersive social experience with co-existence, in the long-lasting social spaces  \cite{ref11}. As anticipated by Gartner \cite{ref2}, over 2 billion people will spend at least one hour a day engaging in social activities in the metaverse by 2026. 
Another report of Meta \cite{ref3} shows that 1.96 billion users daily interacted via Meta in 2022, and every user maintained 338 friendships on average.
Recently, social metaverse has received surge interests on a worldwide scale, driven by the rising demands and the potential for its construction. Besides, many tech giants such as Meta, Roblox, Microsoft, and Epic Games have started to explore the revolutionary new era of social metaverse.

To empower various intelligent social services (e.g., virtual meeting, social dating, and VR live broadcast) with immersive user/avatar interaction in the social metaverse, a vast amount of fine-grained individual data, which can be private and sensitive, is required to be collected and analysed. For example, the built-in cameras and motion sensors in a Oculus helmet can track our head direction, hand/eye movements, facial expressions, and other biometric information to facilitate real-time and immersive user/avatar interaction when we enter the Roblox \cite{ref1}. Besides, in the conventional artificial intelligence (AI) paradigm, these pervasively collected data from geographically distributed individuals is required to be mitigated to a central storage for data analysis, raising severe data misuse and privacy issues.

The emerging federated learning (FL) \cite{ref4}, as a privacy-preserving AI paradigm, is advocated to explicitly preserve the privacy of users/avatars, with the idea of ``bringing the code to the data'' rather than the converse. Still, recent researches \cite{ref5,ref6} validate that user/avatar's privacy can be implicitly divulged under FL in sharing sub-AI models trained from local datasets. Existing countermeasures are generally built on differential privacy (DP) mechanisms for their ease of deployment and low overhead \cite{ref4}. However, the large DP noise to enforce strong privacy provisions also entails a considerable model performance degradation, which eventually reduces the users' quality of experience (QoE). Moreover, the current FL paradigm relies on the central aggregation server for global model aggregation and distribution, whose malfunction can result in a single point of failure (SPoF). Furthermore, the trained AI model under FL is usually valuable and collectively owned by all its trainers, making the authentication of shared ownership to prevent model thefts become a challenge. To sum up, the following three main challenges need to be resolved under FL-empowered social metaverse: 1) efficient privacy-utility tradeoff; 2) robust federated AI (FedAI) model aggregation; 3) ownership provenance of shared FedAI model.

In this article, we propose an efficient and robust FL scheme with ownership provenance and privacy preservation functions for social metaverse applications. 
Specifically, we first leverage the widespread and long-term social connections among users/avatars to advance a social-aware hierarchical FL framework named \emph{SocialFL} for an improved privacy-utility tradeoff in the social metaverse.
Afterward, we design a decentralized SocialFL mechanism based on the blockchain technology, and devise new block structure and consensus protocol based on the on/off-chain collaboration.
Furthermore, based on smart contracts and digital watermarks, we develop an automatic FedAI model ownership provenance mechanism to prevent AI model thefts and collusive users in the social metaverse.

The remainder of this article is organized as follows. The background of social metaverse and the architecture of SocialFL are introduced in Section II. The key challenges of applying SocialFL in the social metaverse are discussed in Section III, and the potential solutions to the social metaverse are presented in Section IV. A case study is given in Section V. The future research directions are offered in Section VI. Lastly, concluding remarks are given in Section VII.

\section{Architecture of SocialFL in Social Metaverse}
\subsection{Overview of Social Metaverse}
\begin{figure}[t]
\centering
  \fbox{\includegraphics[width=9.8cm]{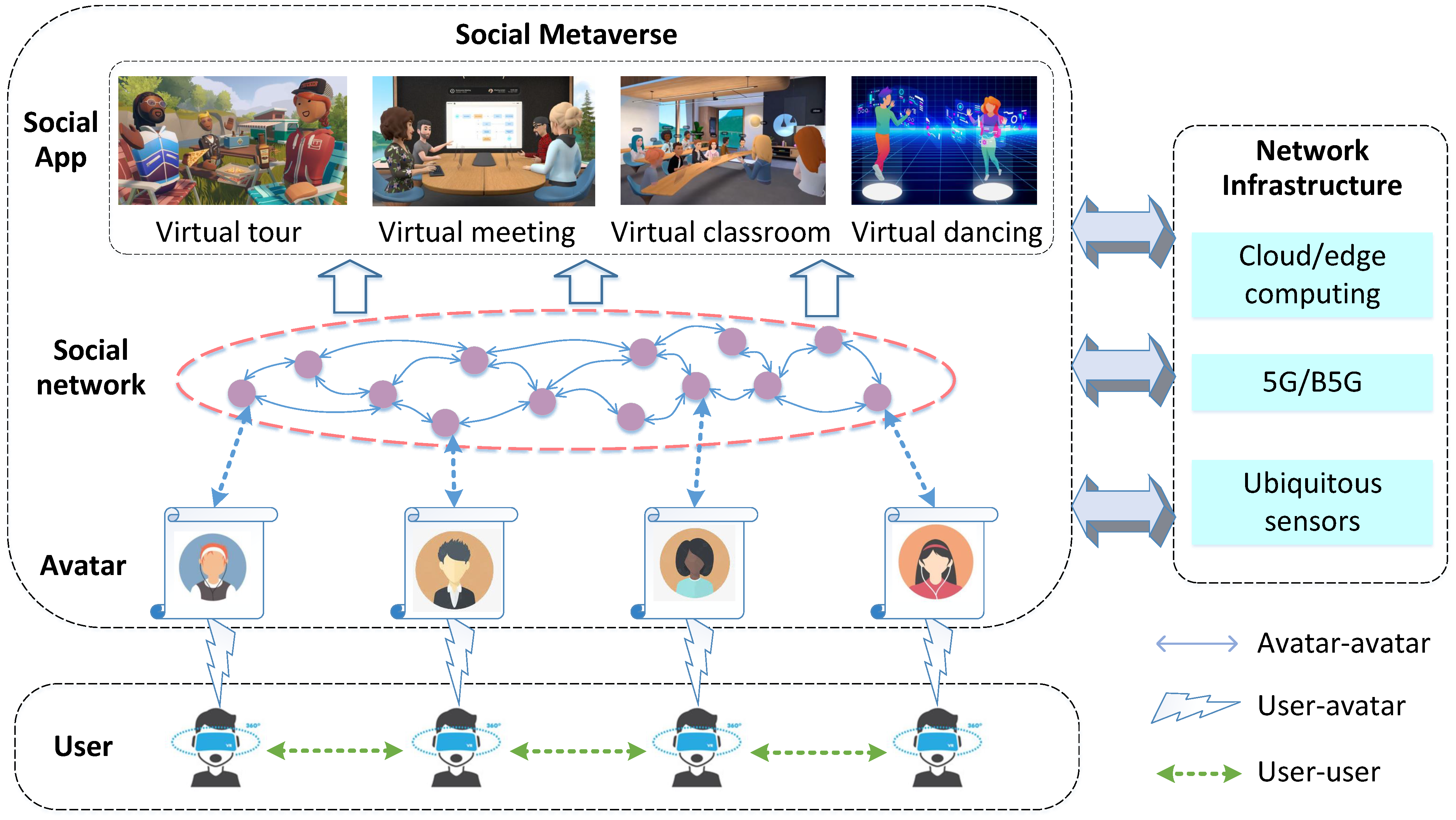}}
  \caption{Overview of the social metaverse.}\label{fig:1}\vspace{-3mm}
\end{figure}

The term ``Metaverse'' was coined by Neal Stephenson in his novel ``Snow Crash'' in 1992, which is described as a computer-generated space fusing digital and physical realities.
Driven by the recent advances in various emerging technologies including AI, blockchain and extended reality (XR), the metaverse is emerged as an exciting upcoming reality in the near future and is expected to revolutionize the current social interactions, social experience, and social commerce. The social metaverse is a shared virtual realm that combines a series of interconnected virtual worlds (also known as sub-metaverses), in which users can play, shop, work, and socialize \cite{ref1,ref11}.
As shown in Fig.~\ref{fig:1}, the social metaverse is composed of user-controlled avatars, digital things, and virtual goods, as below.
\begin{itemize}
  \item \emph{Digital avatars.} Avatars are the virtual embodiments of human users in the metaverse. Users can maintain various avatars in diverse social metaverse applications, and the created digital avatars can be human-like, animals, aliens, monsters, and other imaginary creatures.
  \item \emph{Digital things.} Digital things (e.g., virtual building and digital non-player character (NPC)) in the social metaverse constitute the interactive virtual environments, which can be hyper spatiotemporal (e.g., in ancient or future worlds). 
  \item \emph{Virtual goods.} Virtual goods are the tradeable commodities such as land parcels, skins, digital arts, whose value can be identified by the non-fungible token (NFT) to create a metaverse economic system.
\end{itemize}

All the above elements in social metaverse can be created by users, AI algorithms, or virtual service providers (VSPs). The former refers to the user-generated content (UGC), the middle refers to the AI-generated content (AIGC), while the latter refers to the professional-generated content (PGC). As the social metaverse is parallel and interactive to the real world, there exist the following three kinds of social interactions.
\begin{itemize}
  \item \emph{User-user interactions.} Human users along with their social interactions and social ties (e.g., classmates, friends, relatives, and strangers) constitute the human society.
  \item \emph{User-avatar interactions.} Through XR and human-computer interaction (HCI) technologies, human users can immersively interact with their avatars by wearing XR helmets, smart glasses, etc.
  \item \emph{Avatar-avatar interactions.} Avatars can work, play, collaborate, and socialize with others (e.g., virtual dating, virtual tour, and virtual meeting) in the social metaverse.
\end{itemize}

The social metaverse exhibits unique characteristics from the following perspectives.
\begin{itemize}
  \item \emph{Immersive social experience with co-existence.} In social metaverse, social activities are not through text, image, or video call, but are virtually in-person, bringing the immersion and co-presence that traditional social platforms lack. By capturing users' manifold dimensions and creating true-to-life replicas of physical entities, users/avatars can enjoy immersive sensory experience indistinguishable from the real world.
  \item \emph{Massive and intensive social interactions.} The ever-expanding scale of avatars and UGCs/AIGCs contribute to massive and intensive interactive scenarios in the social metaverse. For example, avatars can hang out with friends in virtual parks or furnish and decorate their virtual houses,  where all types of interactions can co-occur in a shared virtual space. Besides, myriad virtual items and assets can be involved during social interactions.
  \item \emph{Co-created and hyper spatiotemporal social spaces.} Avatars are not only content viewers on a screen, but also digital creators of massive UGCs, contributors of metaverse tools, and participants for metaverse governance. In addition, the social metaverse is free from gravitational, spatial limits, and real life pressures.
  \item \emph{Open and long-lasting social ecosystem.} The social metaverse is an open space for any user with Internet connectivity for various social activities that are previously performed in isolation. Besides, as the social metaverse is physically persistent and not controlled by a single vendor, users' digital identities, social connections, experiences and assets can be long-lasting, thereby promoting a consistent virtual economy system and an independent value system that operate independently of any vendor.
\end{itemize}

\subsection{Social-Aware Hierarchical Federated Learning}
\begin{figure*}[t]
\centering
  \fbox{\includegraphics[width=14.5cm]{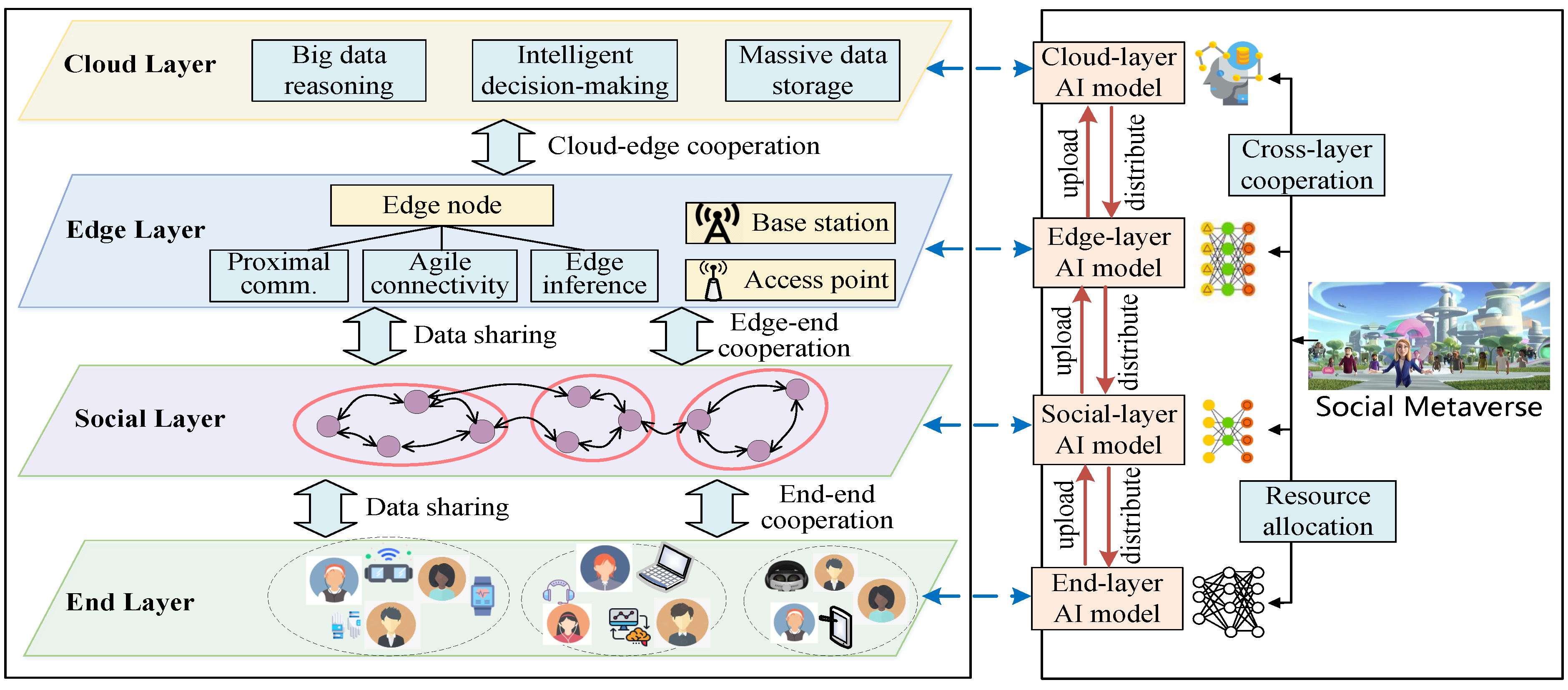}}
  \caption{Architecture of SocialFL with cloud-edge-end orchestration in social metaverse.}\label{fig:2}
\end{figure*}

To empower social metaverse applications using the siloed data on individuals with privacy preservation, as illustrated in Fig.~\ref{fig:2}, the novel SocialFL framework is presented based on cloud-edge-end orchestration, including the following layers.
\begin{itemize}
  \item \emph{On-device AI.} To enhance user QoE, massive heterogeneous XR devices, connected vehicles, and other smart terminals can train on local data in a distributed manner to enable on-device AI services such as collaborative computing and UGC caching. As the AI service occurs directly on terminal devices, the bandwidth consumption can be significantly reduced. Besides, only the local model parameters learned from the data, rather than the original private data, need to be transmitted to upper-layer edge/cloud nodes, thereby explicitly protecting user privacy.
  \item \emph{Social-layer AI.} Considering the ubiquitous social relations among users/avatars, mutually trusted users/avatars can form a social cluster and pre-aggregate their raw local model parameters (instead of the perturbed version injected with DP noises) within the cluster, thereby significantly enhancing model utility. Meanwhile, given the social-layer model aggregation results, it is challenging for other participants and external attackers to deduce the original local model parameters of each member, thereby preventing implicit privacy leaks.
  \item \emph{Edge-layer AI.} Two types of edge nodes: (i) fixed (e.g. base stations and IoT gateways) and (ii) mobile (e.g. vehicles and drones) are considered. By leveraging the distributed edge nodes operated by various VSPs, it realizes edge-layer aggregation of social-layer AI sub-models and facilitates data relays to the cloud for global model aggregation, thereby shortening service latency and improving the quality of FL services. Besides, edge AI capabilities such as proximal communication, edge inference, and agile connectivity can be promoted \cite{ref7}.
  \item \emph{Cloud-layer AI.} The cloud can coordinate the FL process between the edge-layer and the device-layer. Using the powerful cloud computing and storage capabilities, it can efficiently offload the computing and storage services on the user and edge side, and provide users with big data storage and analysis, sophisticated model reasoning, disaster recovery, and other AI services.
\end{itemize}

The social trust between users/avatars is evaluated from the following two aspects.
\begin{itemize}
  \item \emph{Direct social trust.} Direct social trust is built on the direct social interactions between users/avatars, which can be seen as the edge (i.e., social tie) in the social graph. It is related to the experience and duration in each interaction as well as the time decay effect.
  \item \emph{Indirect social trust.} Indirect social trust refers to the topological social trust, which is evaluated based on recommendations from ``friends of friends''. As the user trust can be transferable, the indirect trust is correlated to the direct trust values of all adjacent users on the shortest path connecting them.
\end{itemize}

Users/avatars' social trust is a combination of them. Besides, the social impact of a node in a social cluster is the weighted sum of the number of all neighbor nodes connected to it.

\section{Key Challenges of SocialFL in Social Metaverse}
Driven by increasing demands for effective, robust, and traceable FedAI services, this section highlights the following new challenges arisen in the social metaverse.

\subsection{Stable Social Cluster Formulation}
Current DP-based FL schemes usually require a tradeoff between privacy protection and model utility \cite{ref3}. Existing researches mainly focus on the optimization of FL algorithms, while ignoring the impacts of intrinsic connections between users/avatars such as social ties. By exploiting the social attributes of users/avatars, SocialFL enforces stronger privacy provisions together with enhanced model performance, as mutually trusted users/avatars can form disjoint social groups and aggregate their original local updates inside each group.

However, in social metaverse, users/avatars usually have diverse social connections and complex competition/cooperation features. Moreover, due to the selfishness and profit-seeking nature of users/avatars, they can determine whether to join a social cluster or form a singleton by adopting the solo-training strategy. Thereby, how to form stable and optimal social clustering structures in social metaverse requires further investigation.
Besides, due to the heterogeneity of data size, data quality, and data distribution among metaverse users, how to equitably distribute benefits within each social cluster while prohibiting free-riding behaviors remains an open issue.

\subsection{Robust FedAI Model Aggregation}
Traditional FL paradigm generally relies on a central aggregator, which can result in performance bottlenecks and is prone to single point of failure (SPoF) \cite{ref8}. 
Besides, due to the heterogeneity of metaverse terminals, intermittent wireless communications, and low-latency service requirements in the social metaverse, it leads to low data sharing and model training efficiency in the current decentralized FL paradigm. 1) Due to the significant heterogeneity of metaverse terminals and their limited computing, communication and storage capabilities, they cannot withstand the fast-growing burden of massive data storage and intensive computations in traditional consensus process. Hence, it necessities an efficient and robust blockchain consensus protocol with on/off-chain collaboration and cloud-edge-end collaboration. 2) Since the blockchain technology was initially intended for point-to-point (P2P) transactions, it remains a challenging issue to combine the open blockchain ledgers with the privacy-enhanced SocialFL framework to design new transaction structures and block structures suitable for social metaverse applications.

\subsection{Verifiability of Shared FedAI Ownership}
Generally, training a qualified AI model requires enormous computation power investment, yet the trained AI modes can be vulnerable to model extraction and stealing attacks (e.g., via model compression techniques). Digital watermark technology \cite{ref9} advocates an effective and practical approach to verify the ownership of AI models, in which the model owner embeds its private watermarks as trigger sets for ownership authentication under disputes.

However, existing digital watermark-based solutions for AI model provenance only involves a single model owner, while the shared ownership of AI model in FL settings is rarely considered \cite{ref10}.
1) Due to the diversity and concealment of infringing behaviors, it is difficult to design a secure shared ownership verification mechanism under FL with all-in-round defensive capabilities. 
For example, malicious insiders can illegally resell the shared AI model for profits without the consent of all the owners. They may also steal, forge, and remove the joint watermark via collusion attacks and ambiguous attacks \cite{ref10}.
2) Practical FL services are usually accompanied by the dynamic user joining and exiting process. Besides, participants' local training datasets can be dynamically updated over time. To frequently update the joint watermark, it typically requires a time-consuming and costly model retraining process under traditional mechanisms. How to produce self-evolving joint watermarks under FL with minimal overheads remains a challenge. 

\section{Solutions to SocialFL in Social Metaverse}
This section investigates the game-theoretical stable social cluster formulation (in Sect.~\ref{subsec:scheme1}), lightweight blockchain-based decentralized FL (in Sect.~\ref{subsec:scheme2}), and watermark-based shared FedAI ownership verification (in Sect.~\ref{subsec:scheme3}).
\subsection{Game-Theoretical Stable Social Cluster Formulation}\label{subsec:scheme1}
In this subsection, we design a game-theoretical approach to form a Nash-stable social cluster structure in the social metaverse. Specifically, for each social cluster, its federated payoff is the difference between the utility of the social-layer pre-aggregation model and the federated cost. Here, the model utility is evaluated by its quality (e.g., accuracy).
The federated cost is linearly related to the number of avatars in the social cluster. Within each social cluster, to ensure revenue fairness, the individual payoff of each avatar is the sum of its non-cooperative payoff and the assigned extra payoff. Here, the non-cooperative payoff means the payoff when the avatar forms a singleton (i.e., non-cooperative with others). The extra payoff is the difference between the federated payoff and the sum of non-cooperative payoffs of all the members. Each member receives a portion of the extra payoff according to their learning contributions (i.e., the quality of avatars' local model updates). 

Given the current social cluster structure, each avatar faces three choices: (i) stay in the current cluster; (ii) leave the current cluster and join another non-empty cluster; (iii) detach from the current cluster and form a singleton. The former two are clustered training strategies, and the latter is solo-training strategy. The proposed game-based stable social group formation algorithm includes the following three steps. 1) According to the individual payoff, each avatar constructs a preference order for all social clusters that have not rejected it.
Then each avatar requests to transfer to the cluster (including the empty set) with the largest preference order.
The avatar directly leaves the original cluster and forms a singleton if the most desired option is an empty set. If there exists no transferable clusters even the empty set, it stays in the current cluster.
2) For each avatar that requests to join, the social cluster admits it as a candidate if the payoff of any member within the social cluster can be improved when the avatar joins in. The social cluster builds a preference order for all candidates, only admits the one with the highest preference order, and rejects others.
3) Repeat the above process until no avatar tends to leave the current cluster to seek better individual returns (i.e., reaching the Nash-stable state).

\subsection{Lightweight Blockchain-Based Decentralized SocialFL}\label{subsec:scheme2}
\begin{figure}[t]
\centering
  \fbox{\includegraphics[width=9.5cm]{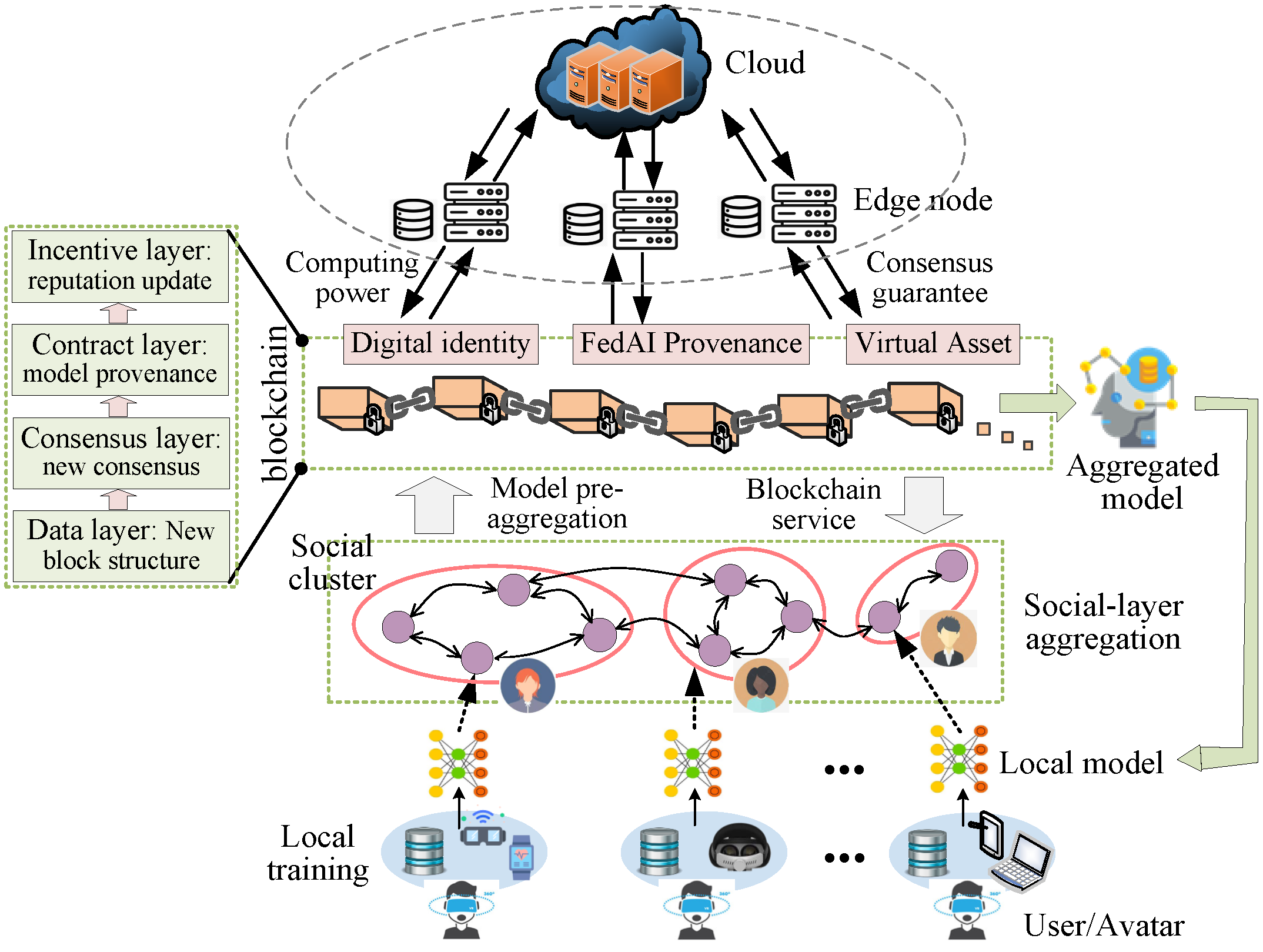}}
  \caption{Blockchain-enabled decentralized SocialFL in social metaverse.}\label{fig:3}
\end{figure}

In this subsection, we devise a lightweight blockchain-based approach for robust and decentralized AI model aggregation in the social metaverse. As shown in Fig.~\ref{fig:3}, three types of blockchain nodes are considered: full node, light node, and validator node. Edge nodes and cloud servers serve as full nodes who store all historical block information; while metaverse terminals act as light nodes who only store the block header and obtain blockchain services from full nodes. Validator nodes are elected from full nodes, and they are responsible for blockchain consensus maintenance and ledger management.
To adapt to the FL setting and improve model aggregation efficiency, we design a new block structure named \emph{model aggregation (MA) block} in each global communication round under FL. In particular, we add the metadata including FL task ID, global aggregation results, a seed (for validator election), and a consensus voting script (CVScript) into the header of the MA block.
In the block body, the following three new transaction types are designed.
\begin{itemize}
  \item \emph{FL task request transaction (trTx)}, which records the FL task-related information including the task ID, timestamp, signature, task reward, initial AI model parameters, and expected AI performance index.
  \item \emph{Social-layer model aggregation transaction (saTx)}, which includes pre-aggregated model parameters within a social cluster, members' contributions, and the multi-signature of all the members.
  \item \emph{Global model aggregation transaction (gaTx)}, which includes the global aggregation results conducted by any consensus node as well as its signature.
\end{itemize}

\begin{figure}[t]
\centering
  \fbox{\includegraphics[width=9.5cm]{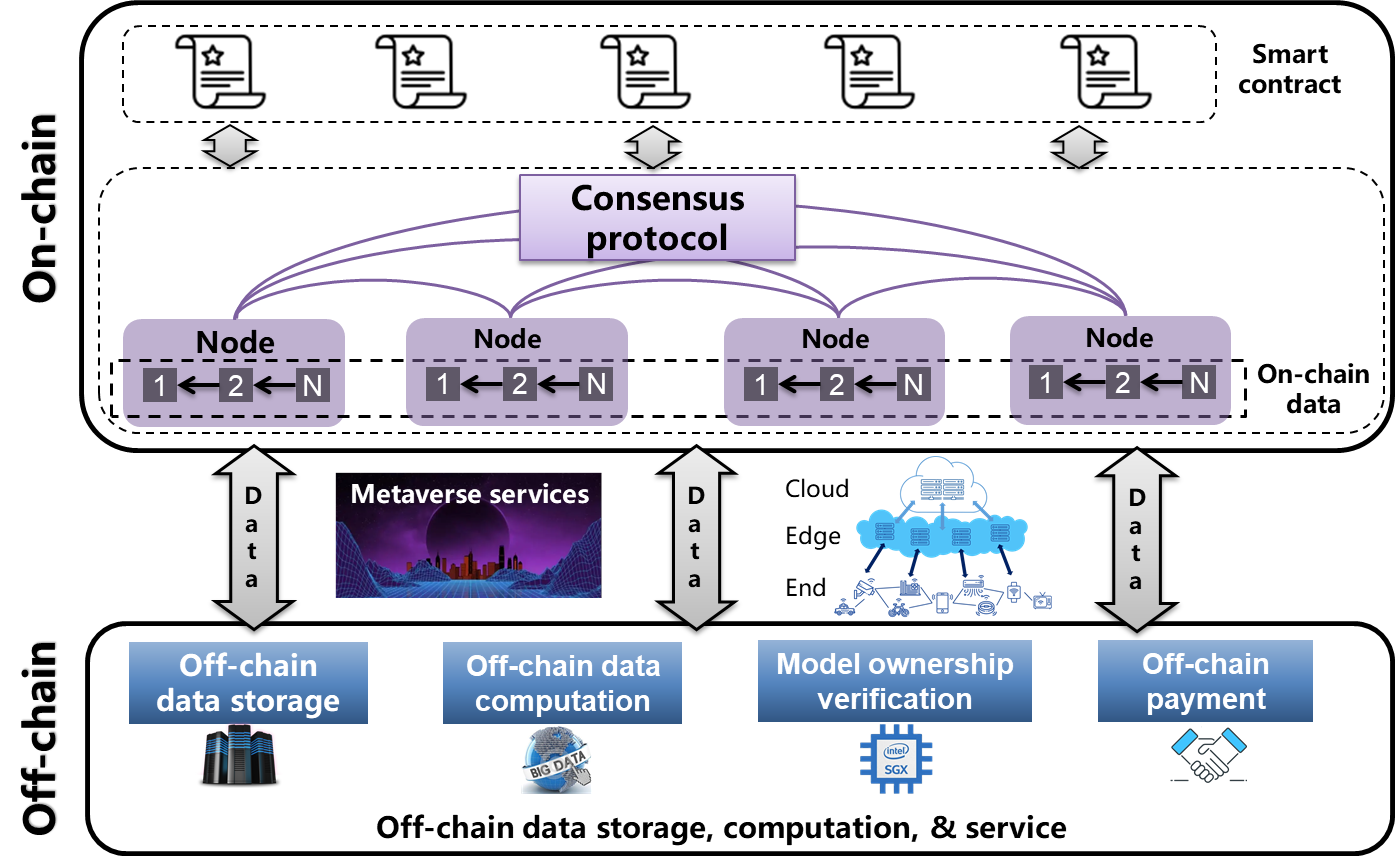}}
  \caption{Illustration of on/off-chain collaboration in blockchain service offering.}\label{fig:4}
\end{figure}

As shown in Fig.~\ref{fig:4}, we devise an improved consensus protocol based on on-chain and off-chain collaboration for enhanced consensus efficiency from the perspectives of storage, computing, payment, and service. (1) \emph{Storage}: the CVScript, social-layer pre-aggregated model parameters, and global model parameters are stored in the off-chain data warehouses, while only the corresponding hash pointers are retained on the blockchain to lower on-chain storage costs; (2) \emph{Computation}: validator nodes compute nodes' reputation values based on the CVScript in an off-chain manner and run the consensus protocol to update the on-chain reputation status of nodes for efficient reputation update; (3) \emph{Payment}: the VSP builds its hashchain and sequentially delivers the hash value in the hashchain as an off-chain payment commitment, while the final payment is completed on the chain, thereby reducing the transaction overheads;
(4) \emph{Service}: the trusted execution environment (TEE) executes the off-chain model ownership verification and provenance operations, and the smart contract synchronizes the provenance results on the blockchain.
The proposed consensus protocol includes the following four phases.

\textbf{Phase 1: Consensus committee formation.} The validator committee is elected via the verifiable random function, and the chance that each full node to be elected is determined by its reputation value.

\textbf{Phase 2: Candidate block proposal and dissemination}. At the initial stage of each block height, each validator independently verifies new transactions and package the successfully verified ones into a candidate MA block proposal.

\textbf{Phase 3: Multi-stage consensus voting}. At each stage, members of the consensus committee are dynamically updated to prevent from being targeted. Validators carry out the Algorand Byzantine agreement method to reach consensus through multiple rounds of consensus voting to ensure that the outcome of each consensus round is witnessed by enough validators to ensure consensus security.

\textbf{Phase 4: Add a new block}. After reaching consensus, a newly built MA block is linearly appended to the blockchain, which contains a hash to link its parent block. The CVScript is the aggregation of votes signed by validators nodes in multi-stage voting, which allows new validators to catch up with the new block verification process. Based on CVScript, the reputation of honest validators is increased by $\Delta_1$ as a reward, while the reputation of validators that propose false blocks is decreased by $\Delta_2$ as a punishment.

\subsection{Watermark-Based Shared FedAI Ownership Verification}\label{subsec:scheme3}
\begin{figure*}[t]
\centering
  \fbox{\includegraphics[width=13.5cm]{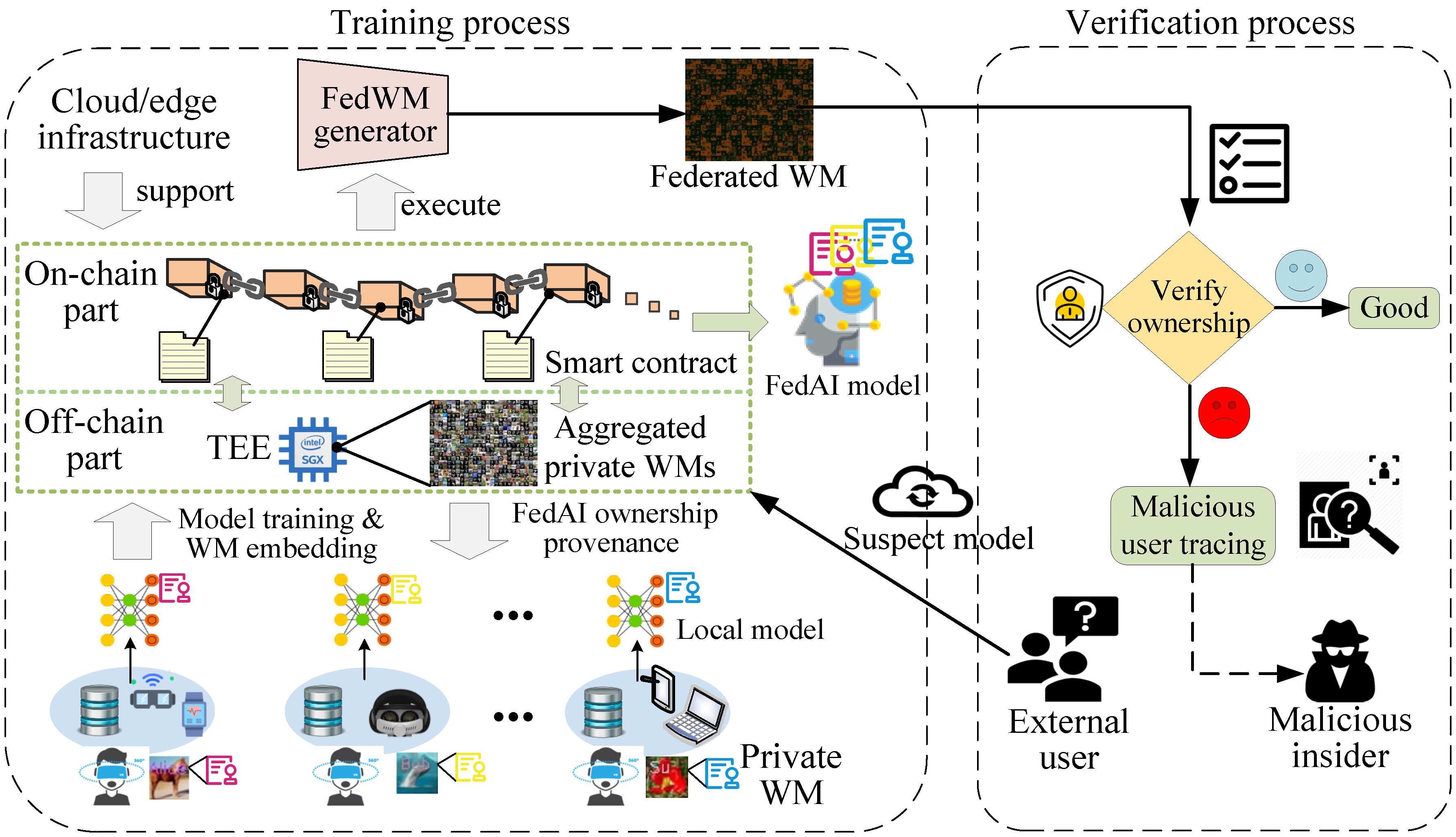}}
  \caption{Digital WM-based shared FedAI model ownership verification in social metaverse.}\label{fig:5}
\end{figure*}

This subsection introduces a practical and adaptive watermark-based approach for FedAI intellectual property protection in FL scenarios, where all clients equally share the ownership of the well-trained FedAI model and any disposal of the model has to be approved by all owners. It contains the following two phases.

\textbf{Phase 1: Generation and Embedding of the Joint Watermarks:}
To defense and trace internal infringers, we produce the joint watermark as a fingerprint of the FedAI model. Firstly, in local training, each client trains its local model using its private data and private trigger set, i.e., a group of random images tagged with pre-designed (always wrong) labels. After certain rounds of local training, each client's model weights embedded with its private watermark are uploaded to the aggregation servers (e.g., edge nodes and the cloud) for global aggregation based on the blockchain. Moreover, their private watermarks are uploaded to the off-chain TEE (e.g., Intel SGX) without exposure to any other clients. Here, the TEE is assumed to be secure. Specifically, the TEE trains a generative adversarial network (GAN) to fuse the uploaded private trigger sets and produce a joint watermark. To prevent internal infringements, the soft label, generated from the normalization of all hard labels of private trigger sets, serves as the label of the joint watermark. Each aggregation server then independently aggregates the local weights into a global model and fine-tunes the global model by embedding the joint watermark with soft label. Next, the consensus global model can be produced via the consensus process and distributed to every client for next-round training.

\textbf{Phase 2: Authentication and Verification:}
For any suspect FedAI model, an external user can invoke the ownership provenance smart contract for ownership verification. As the ownership of FedAI model is collectively shared, any external verification has to be approved by all owners by sending their own private watermarks to the TEE to grant authorization. As malicious owners may collude to forge others' private watermarks, a scoring mechanism is devised based on the diversity and similarity to assure that each client has uploaded its private watermarks. 
If the score of the uploaded private watermarks exceeds a preset threshold, the TEE produces a joint watermark using GAN and inputs the joint watermark into the suspected model via the remote API. If the gap between the output and the soft label is below a preset small threshold, the infringement behavior of FedAI model can be identified with overwhelming probability to enforce accountability. 

\section{Case Study}
\begin{figure*}[!t]
\fbox{\centering
\subfigure[]{
\label{fig:6-1}
\includegraphics[width=7.5cm]{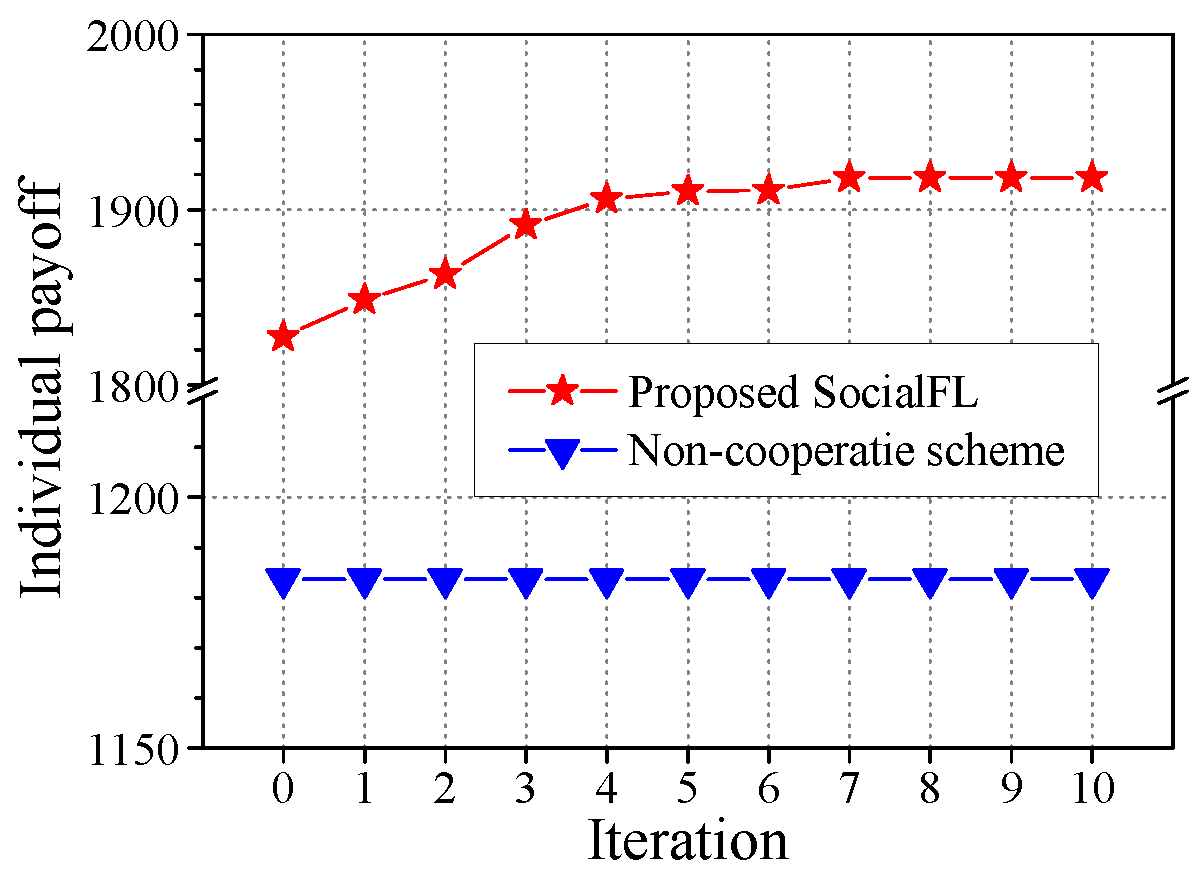}}\ \ \ \
\subfigure[]{
\label{fig:6-2}
\includegraphics[width=7.5cm]{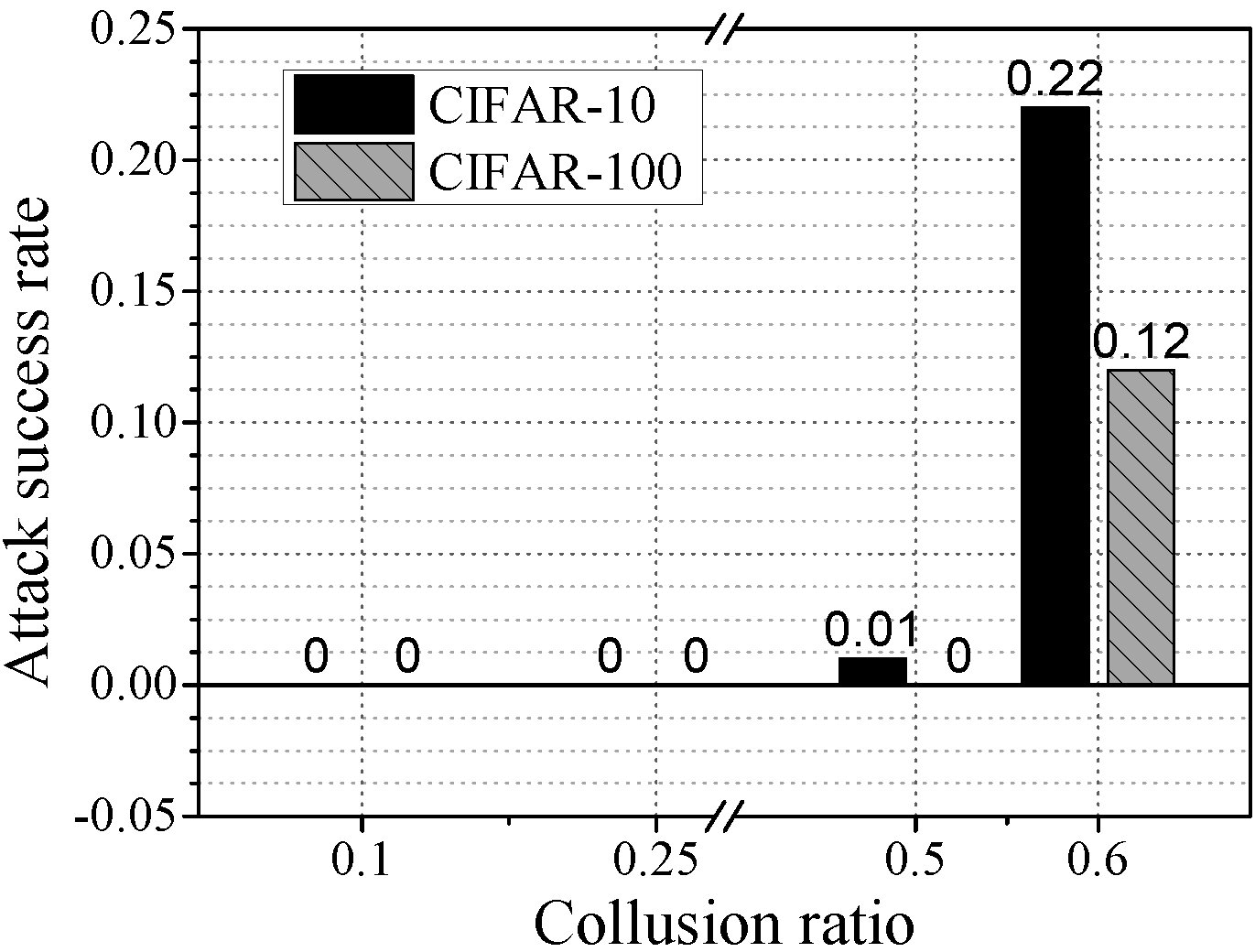}}\ \ }
\caption{a) Evolution of average individual payoff over time in SocialFL compared with the non-cooperative scheme.
b) Attack success rate under different collusion ratios and datasets in SocialFL.}
\label{fig:6}
\end{figure*}

In the case study, we employ PyTorch to implement the SocialFL. We consider 100 participants, and their direct social ties in the social graph follow the random distribution between [0, 1]. Two image recognition FL tasks using the CIFAR-10 and CIFAR-100 datasets are considered. Each dataset is partitioned among the participants in the non-IID manner controlled by the Dirichlet distribution with concentration parameter of 10. Each participant adopts ResNet18 for local training with SGD learning rate of 0.01 and batch size of 64. Each participant randomly opts 100 images with bogus labels as the trigger set, and uploads 10 private watermarks to the TEE to produce a shared watermark. Two collusion attacks, i.e., joint watermark stealing and joint watermark counterfeiting, are considered in the simulation.

In the non-cooperative scheme, each user behaves uncooperatively and adds moderate DP noises into local model updates. As seen in Fig.~\ref{fig:6-1}, the average individual payoff fast converges after about 7 iterations in our SocialFL scheme and achieves a significantly better individual return for participants than the non-cooperative scheme. In addition, as seen in Fig.~\ref{fig:6-2}, our proposed scheme can effectively resist collusion attacks during FedAI ownership provenance for both two datasets when the collusion ratio is less than 50\%. 

\section{Future Research Directions}
In this section, the open research directions are discussed including but not limited to the following.
\subsection{Social-Aware Personalized Learning in Metaverse}
Metaverse users usually have diverse privacy tastes in performing different FL tasks. 
Besides, the datasets among distributed individuals are generally heterogeneous, namely, non-independent identically distributed (non-IID). Personalized FL (PFL) \cite{ref12} is advocated to resolve the fundamental issues of FL with personalized privacy provisions and heterogeneous data. In PFL, instead of sharing one global model and the same privacy provision for all clients, multiple produced global models can be produced and each of them can adapt better with the user's own data distribution while enforcing personalized privacy protection by adding DP noises scaled to its privacy preferences. For improved learning efficiency in metaverse, the social links between users can be utilized to group users with similar interests and social trusts to produce customized DP noises for users with varying privacy levels.

\subsection{Orchestration of Blockchain and AI in Metaverse}
On one hand, the disruptive AI technology brings numerous benefits to the blockchain-enabled metaverse such as dynamic smart contract audit, on-chain data analysis, misbehavior provenance, and energy-efficient consensus design \cite{ref13}. On the other hand, blockchain offers trust-free ledgers to facilitate collaborative learning, knowledge sharing, contribution recording, and fair payment among distrustful avatars in the metaverse. The orchestration of blockchain and AI in metaverse needs further investigations \cite{ref14}.
For example, it remains a challenge to design AI-inspired blockchains to automatically audit the known/unknown code vulnerabilities in advance and optimize key parameters such as block size and type of consensus mechanisms for enhanced liveness, scalability and resilience.

\subsection{Semantic Communication Enabled Social Metaverse}
In contrast to conventional communications which focus on accurately transmitting symbols or bit sequences, in semantic communication-enabled social metaverse, only necessary semantic information (i.e., the meaning of message) relevant to the particular social mission is intelligently transmitted from the source to the receiver \cite{ref15}, thereby greatly alleviating data traffic in social metaverse applications. Essentially, AI technologies enable semantic detection, knowledge modeling, semantic encoding/decoding, and avatar coordination in the metaverse. However, new security threats such as semantic data/knowledge poisoning and unintentional privacy violation such semantic privacy leakage can breed. As existing defenses against these threats have not kept pace, it raises necessitates for new defense paradigms for enhanced security and privacy in the complex social metaverse.

\section{Conclusion}
In this article, we have investigated the efficient, robust, and traceable FedAI model training for privacy-preserving social metaverse applications. Specifically, by taking advantage of the mutual social relations between users/avatars, we have presented a social-aware hierarchical FL framework named \emph{SocialFL} to enhance model utility without compromising privacy.
Secondly, we have designed a blockchain-enabled aggregator-free SocialFL mechanism with improved robustness and reliability, as well as new block structures and a reputation-based consensus protocol based on the on/off-chain collaboration.
In addition, an automatic FedAI ownership provenance mechanism has been developed based on smart contracts and digital watermarks to prevent AI model thefts and collusion attacks in the social metaverse.
We have also provided a case study to validate the effectiveness of the proposed framework. Finally, open research issues essential for social metaverse are discussed.


\end{document}